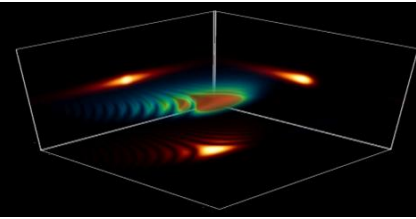

# arXiv Preprint

# Single-shot complete spatiotemporal Strehl ratio metrology of ultrashort laser pulses–theory.

## THOMAS OKSENHENDLER,[1,*]

[1]iTEOX, 14 avenue Jean Jaurès, 91940 Gometz-le-chatel, France
*Corresponding author: thomas@iteox.com

**Ultra-high intensity femtosecond lasers at focus are now routinely used for relativistic motion of charged particles with peak intensities over $10^{18}$W/cm². Such high-field experiments are very sensitive to the value of the peak intensity. Therefore, precise knowledge of the pulse intensity is a key prerequisite for the correct interpretation of experimental data. Accurate measurement of energy, spectral and spatial profiles being well established, peak intensity evaluation then only requires the determination of the spatiotemporal Strehl ratio. Here I propose a new method based on a simple analysis of a nonlinear effect. It diagnoses all the temporal, spatial and spatiotemporal coupling distortions. This method can even be implemented on target by using relativistic surface harmonic generation for example. © 2020 iTEOX**

## I. Introduction

The last three decades have seen the surge of the peak power of ultra-intense lasers up to PW levels (1 PW = $10^{15}$W). Larger facilities aiming at reaching tens of PW are under commission all over the world [1]. For peak intensity higher than the relativistic limit ($10^{18}$ W/cm²), accurate knowledge of the on-target pulse intensity is essential for the correct interpretation of high-field experiments involving highly nonlinear processes. In such high-intensity systems, the value of the on-target intensity is commonly estimated from the spatial and temporal power/fluence distributions, which are measured separately [2, 3]. Such ultra-high peak intensities can only be reached by a combination of temporal and spatial compression. Chirped pulse amplification [4] smartly introduces in the temporal domain a stretching/compressing technology that is simply achieved, in the spatial domain, by focusing. On PW systems, the beam is typically stretched to a volume about 10cm x 10 cm x 1ns (30cm) as illustrated on fig.1.a. The laser pulse is firstly temporally squeezed (fig.1.b) by a grating compressor down to typically 30fs (9μm). It is then spatially focused (fig.1.c) down to diffraction limited spot i.e. with a typical diameter of about 1μm. The total compression factor is in the order of $10^5$ x$10^5$ x3.$10^4 \approx 10^{15}$. Such gigantic compression rate is the only way to reach peak intensities beyond the order of $10^{18}$W/cm².

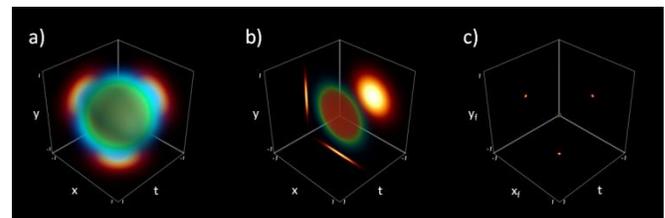

Fig. 1. a) Stretched pulse in X, Y transverse dimension and in time, b) temporally compressed pulse just behind the grating compressor(near field) and c) perfect pulse without distortion focused in space and compressed in time in the focal plane or far field.

While there are reliable approaches for the detection of pulse energy and spatial distribution [5], precise characterization of the on-target temporal structure remains a challenge, especially for intense, few-cycle pulses. Achieving reliable detection of the spatial intensity distribution is simpler than achieving that of the temporal distribution, as it can be done directly with a CCD camera in combination with an imaging objective in the near or far field [5] Furthermore, the result is not affected by dispersive optical elements in the beam path (at least when they are of good enough quality and the beam is properly attenuated to prevent nonlinear distortions). One should note that a reliable measurement requires a high dynamic range (HDR) image to detect the energy around the main peak (spatial background). This one, in the far field, is due to high frequency amplitude or phase modulation in the near field. It can be measured with a combination of an HDR CCD camera whose dynamic range is extended by the sequential measurement of unsaturated and saturated images corresponding to different attenuation factors on the beam [6]. It can alternatively be measured in the near field where the required dynamic is substantially decreased by the compression rate. Nevertheless, it is then a matter of having a high enough resolution in amplitude and phase to catch the high frequency components responsible for the background. The ultimate limit in resolution that matters is the optical wavelength of the pulse, i.e. about 1μm. Consequently, the beam should ideally be measured in the near field with a 1μm resolution.

In the temporal domain ultra-high peak intensity requires higher dynamic characterization to avoid any pre-ionization at levels lower than $10^{12}$W/cm². Specific ultra-high dynamic contrast measurements can assess the quality of the contrast in single-shot [7, 8].

Even with a full high dynamic spatial and temporal characterization, such beams, due to their large size and ultra-high compression rate can have spatiotemporal couplings (STC) that degrade the spatiotemporal shape of the pulse on target [9–13]. CPA is known to introduce such STC through both the stretcher and compressor when not perfectly aligned or due to grating defects [9, 12, 13]. The most common STC are pulse-front tilt (PFT) [14], transverse group delay dispersion (TGDD) [15] and pulse-front curvature (PFC) [8], illustrated in near and far field on figure 2. Each coupling corresponds to a specific distortion between the pulse front and the phase front generated through specific optics (prisms, gratings, lenses). Such STC dramatically reduces the peak power available on target, making the control of the spatiotemporal shape of one's laser mandatory, in addition to control of the temporal and spatial profiles.

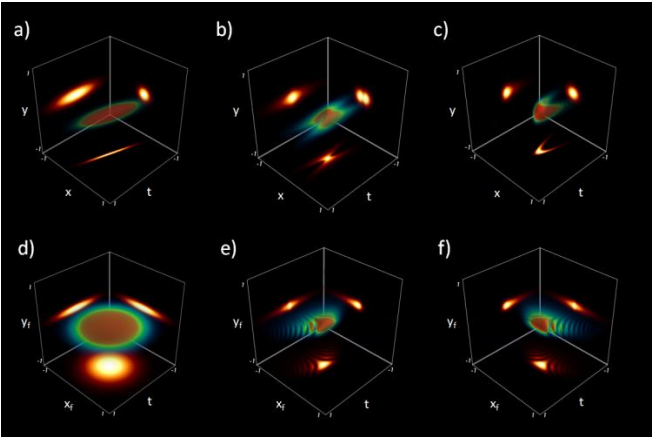

Fig. 2. Pulses with a) PFT, b) TGDD and c) PFC in near field, and in far field respectively (d,e,f).

As shown on figure 2, the temporal measurement at one position in the near field cannot allow the detection of the influence of STC on the global pulse duration.

The full characterization of such laser pulses in a particular space-time domain requires the knowledge of both the amplitude and the phase of the electric field (in this domain) [16]. With an imaging spectrometer, we have a direct single-shot access to the amplitude in the near field spatiospectral domain $(x, \omega)$ or the far field $(x_f, \omega)$ spatiospectral [17] ($x$ is one transverse spatial dimension, $x_f$ its spatial far field proportional to a spatial frequency). As usual in ultrafast pulse metrology, the difficulty arises from the determination of the phase. A Fourier transform of the product of this amplitude and its associated spatiospectral phase yields the spatiotemporal shape of the pulse. Since the imaging spectrometer only provides information about one transverse spatial dimension, full single-shot characterization will require at least two simultaneous measurements over the two transverse spatial dimensions $x$ and $y$. Considering that the complete signal to recover is $A(x, y, \omega)e^{i\varphi(x,y,\omega)}$ on ranges (10cm, 10cm, 1ns) with resolutions (1µm, 1µm, 10fs), the amount of data to be measured ($10^{13}$ measurement points) is considerable. The complexity of the means of measurement to be implemented becomes so great that the question of its relevance and practicality arises. Such large laser systems become so complex that the diagnostics applied must be inspired by other fields of complex systems such as for example medicine. Certain parameters make it possible to qualify the patient's good health. What are they for ultrashort ultra-intense pulse lasers?

For relativistic laser-matter interaction at peak laser intensities beyond $10^{18} W/cm^2$, knowledge of the peak intensity is the main parameter. Others are also very important such as the temporal contrast. Peak intensity may be seen as the temperature of the patient in medicine.

Knowing the full energy of the pulse, the peak intensity can be estimated by determination of the pulse amplitude in the near field and spectral domain $(x, y, \omega)$. The 3D Fourier transform of this amplitude will then give the best pulse and highest peak intensity that can be obtained. Actual peak intensity determination requires including phase distortions in the initial field.

As for the spatial distribution, one can use the full spatiotemporal Strehl ratio [18-20] hereinafter referred as full Strehl ratio (SR$_{full}$). SR$_{full}$ is defined as the ratio between the maximum of the real peak intensity including all phase distortions and the peak intensity of the pulse corrected from these phase distortions:

$$SR_{full} = \frac{I[A(x,y,\omega)e^{i\varphi(x,y,\omega)}]}{I[A(x,y,\omega)]} \qquad (1)$$

where x, y are near field spatial dimensions, $\omega$ is the optical pulsation ($\omega = \frac{2\pi c}{\lambda}$, c is the speed of light in vacuum and $\lambda$ the optical wavelength of the pulse), A and $\varphi$ are respectively the amplitude and the phase of the electric field in the spatiospectral domain, I[ ] denotes the calculation of the maximum of the focused intensity of a given beam:

$$I[E(x,y,\omega)] = \max\left(\left|TF_{3D}^{-1}[E(x,y,\omega)]\right|^2\right) \qquad (2)$$

This full Strehl ratio is more simply defined in the far field spatiotemporal domain ($x_f, y_f, t$) as:

$$SR_{full} = \frac{Max[I(x_f,y_f,t)]}{Max[I_{TF}(x_f,y_f,t)]} \qquad (3)$$

where $I(x_f, y_f, t)$ is the 3D Fourier transform of $A(x, y, \omega)e^{i\varphi(x,y,\omega)}$ and $I_{TF}(x_f, y_f, t)$ is its Fourier transform limited (FTL) pulse i.e. the 3D Fourier transform of the amplitude without any phase distortions [19]. Knowing SR$_{full}$, the near field spatial and spectral distribution and the pulse energy completely defines the pulse peak intensity. As mentioned before, pulse energy, spatial and spectral profile measurements despite their relative complexities are achievable in single-shot even on complex systems. Spatiospectral amplitude profiles can also be measured in single-shot by using imaging spectrometers [17]. All the complexity of the temporal and spatiotemporal measurement can then be addressed by the estimation of the full Strehl ratio. It includes the temporal, spatial and spatiotemporal intensity distortions [20].

This article focuses on the metrology of the full Strehl ratio to quantify the peak intensity of the ultrafast laser pulse. Although it is written in the context of ultrashort pulse metrology, no direct phase measurement or retrieval will be implemented. The full Strehl ratio meter acts as diagnosing the pulse compression in single-shot. However, the determination of the causes of the distortions requires additional examinations. After an introduction on the effects of phase distortions on the peak intensity and full Strehl ratio, section II presents common situations where spatiotemporal coupling detrimentally affects the pulse. Notions of global and local widths then reveal that the full Strehl ratio is determined by the upper limit of partial or local Strehl ratios combination. Section III introduces a new diagnostic method, Immediate Pulse Optimal Compression Comparison Estimation (IMPULSOCCE), constituting a full Strehl ratio meter. After a brief description of the method, the theory is introduced for any nonlinear effect. Numerical simulations show its efficiency for 2nd and 3rd order distortions. To generalize these first results, Section IV introduces the concept of spectral phase frequency as the spatial phase frequency is used to extend the analysis to any type of distortions. The last section gives examples on determining the origin of the distortion for 2nd and 3rd orders, low and mid phase frequencies. For convenience, the acronyms and the notation used throughout are summarized in Tables 1 and 2.

| Variable or function | Definition |
|---|---|
| STC, CPA, PW, FTL | Spatiotemporal Coupling, Chirped Pulse Amplification, PetaWatt, Fourier Transform Limited |
| RMS, HDR, CCD | Root Mean Square, High Dynamic Range, Coupled Charge Device |
| PFT, TGDD, PFC | Pulse-Front Tilt (PFTx: $\varphi(x,\omega) = ax\omega$), Transverse Group Delay Dispersion (TGDDx: $\varphi(x,\omega) = ax\omega^2$), Pulse Front Curvature (PFCx: $\varphi(x,y,\omega) = ax^2\omega$) |
| OPL | Optical Path Length |
| LSF, MSF, HSF | Low-Spatial Frequency, Mid-Spatial Frequency, High-Spatial Frequency |

TABLE 1. Acronyms summary.

| Variable or function | Definition |
|---|---|
| x, y | Near field spatial dimensions, normalized and centered ( x or y $\in$[-1,1] and <x>=<y>=0) |
| $x_f$, $y_f$ | Far field spatial dimensions, normalized and centered ( $x_f$ or $y_f \in$[-1,1] and <x>=<y>=0 |
| $\omega$ , Frequency, t | Optical pulsation:$\omega = \frac{2\pi c}{\lambda}$, c is the speed of light in vacuum and $\lambda$ the optical wavelength Optical frequency: $Frequency = \frac{\omega}{2\pi}$. Time: t |
| (x, y, $\omega$) | Near field spatiospectral domain normalized and centered (also $\omega \in$[-1,1] and < $\omega$>=0) |
| $(x_f, y_f, t)$ | Far field spatiotemporal domain normalized and centered ( also $t \in$[-1,1] and < t>=0 ) |
| E(x, y , $\omega$) | Analytic electric field in near field spatiospectral domain: $E(x,y,\omega) = A(x,y,\omega)e^{i\varphi(x,y,\omega)}$ |
| A(x, y , $\omega$) | Analytic electric field pulse envelope in near field spatiospectral: $A(x,y,\omega) = \frac{e^{-x^2/2\Delta x^2}}{\sqrt{2\pi}\Delta x}\frac{e^{-y^2/2\Delta y^2}}{\sqrt{2\pi}\Delta y}\frac{e^{-\omega^2/2\Delta\omega^2}}{\sqrt{2\pi}\Delta\omega}$ |
| $\Delta x, \Delta y, \Delta\omega$ | Widths of the pulse in the near field spatiospectral ( $\Delta x$ =0.141,$\Delta y$ =0.173, $\Delta\omega$ =0.1) defined as standard deviation, as example for $\Delta x$: $\Delta x = \left[\frac{\int_{-\infty}^{\infty} x^2 I(x)dx}{\int_{-\infty}^{\infty} I(x)dx} - \left(\frac{\int_{-\infty}^{\infty} xI(x)dx}{\int_{-\infty}^{\infty} I(x)dx}\right)^2\right]^{1/2} = [\langle x^2\rangle - \langle x\rangle^2]^{1/2}$. |
| I(x,y,$\omega$) | Near field spatiospectral intensity: $I(x,y,\omega) = \|A(x,y,\omega)\|^2$ |
| TF$_{1D}$, TF$_{1D}^{-1}$ | Fourier transform and inverse Fourier transform along a single dimension |
| E(x, y, t) | Analytic electric field in near field spatiotemporal domain: $E(x,y,t) = A(x,y,t)e^{i\varphi(x,y,t)} = TF_{1D}[A(x,y,\omega)e^{i\varphi(x,y,\omega)}]$ |
| TF$_{3D}$, TF$_{3D}^{-1}$ | Fourier transform and inverse Fourier transform in the three dimensions |
| $E(x_f, y_f, t)$ | Analytic electric field in far field spatiotemporal domain: $E(x_f,y_f,t) = A(x_f,y_f,t)e^{i\varphi(x_f,y_f,t)} = TF_{3D}[A(x,y,\omega)e^{i\varphi(x,y,\omega)}]$ |
| $I(x_f,y_f,t)$ | Far field spatiotemporal intensity: $I(x_f,y_f,t) = \|A(x_f,y_f,t)\|^2$ |
| $\Delta x_f, \Delta y_f, \Delta t$ | Widths of the pulse in the far field spatiotemporal domain |
| $I_{TF}(x_f,y_f,t)$ | Far field spatiospectral Fourier transform limited pulse intensity: $I_{TF}(x_f,y_f,t) = \|TF_{3D}[A(x,y,\omega)]\|^2$ |
| $\Delta x_{f_0}, \Delta y_{f_0}, \Delta t_0$ | Widths of the Fourier transform limited pulse in the far field spatiotemporal |
| $SR_{full}, SR_{x_f y_f t}$ | Full Strehl ratio: real and ideal Strehl ratio estimated by simulation of the peak intensities, local Strehl ratios are defined by indexing the considered dimensions: $SR_{full} = SR_{x_f y_f t}$ |
| $I_{NL}(x_f,y_f,t)$ | Far field spatiotemporal nonlinear pulse intensity for order N: $I_{NL}(x_f,y_f,t) = [I(x_f,y_f,t)]^N = \|A(x_f,y_f,t)\|^{2N}$ |
| $I_{NL}(x,y,\omega)$ | Near field spatiospectral nonlinear pulse intensity for order N: $I_{NL}(x,y,\omega) = TF^{-1}_{3n}[E(x_f,y_f,t)^{2N}]$ |
| $\Delta x_{NL}, \Delta y_{NL}, \Delta\omega_{NL}$ | Widths of the nonlinear pulse in the near field spatiospectral domain |
| $I_{TFNL}(x,y,\omega)$ | Near field spatiospectral nonlinear Fourier transform limited pulse intensity for order N: $I_{NL}(x,y,\omega) = TF^{-1}_{3D}[\|TF_{3D}[A(x,y,\omega)]\|^{2N}]$ |
| $\Delta x_{0NL}, \Delta y_{0NL}, \Delta\omega_{0NL}$ | Widths of the nonlinear pulse in the near field spatiospectral domain |
| $\Gamma_{kl(ij)}(u,v)$, $\Gamma mes_{kl(ij)}(u,v)$ | Broadening factor along dimensions k,l ( theoretical or measured mes) as a function of variables (u,v), indices <i,j> indicate integrated dimensions. |
| $R_{kl(ij)}(u,v)$, $R_{lmeskl(ij)}(u,v)$ | Estimated local Strehl ratios for dimension k,l by the method as a function of variables (u,v), indices <i,j> indicate integrated dimensions. |
| $R_{full f}, R_{l3D}, R_{l2D}$, $R_{lmes}(u,v), R_{l4\omega}$ | Estimated and measured full Strehl ratios |

TABLE 2. Abridged notation summary. The appended numerical values are those used in the examples.

## II. Influence of phase distortions on peak intensity

Many examples were given through the literature on the phase distortions effects first in the spatial domain [21], then for ultrashort pulses in the temporal domain [22], and finally for ultra-intense ultra-short pulses in the spatiotemporal domain [9-16,17,19,23]. Influence of low frequency spatial phase components on the pulse energy distribution are well documented and can be easily taken into account to estimate peak intensity. High frequency phase distortions are less accessible despite their influence on the peak intensity that are of prime importance [24]. To illustrate these influences on the far field spatiotemporal amplitude, the initial pulse is distorted by low frequency spatial phase (2$^{nd}$ order) (fig.3.a) or by high frequency modulations (fig.3.b) which decrease its peak intensity by a factor of 2.As shown by the spatiotemporal intensity on fig.3.c it is nearly invisible on the intensity as the energy lost is spread over a large volume. This volume is getting larger as the stretched pulse is large. The $10^{15}$ compression factor used for PW level systems is very challenging on the measurement dynamic.

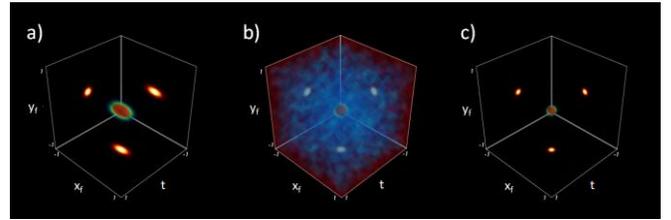

Fig. 3. Far field spatiotemporal amplitudes of pulses with a) pure spatial 2$^{nd}$ order on x dimension with SR=0.5, b) high frequency phase modulation (x,y, ω) with SR=0.5 and c) its far field spatiotemporal intensity.

The high frequency components can be seen through their spatial contribution as they do modulate the spatial amplitude when the pulse is propagating over relatively short distances as illustrated on fig.4. Thus, such high frequency can be measured on amplitude close to near field measurements (fig.4.c).

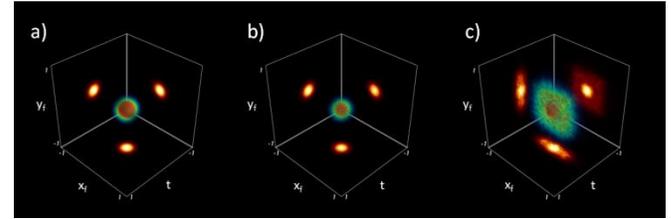

Fig. 4. Spatiotemporal amplitudes of pulses with a) high frequency phase modulation (x,y, ω), b) after propagation over z=10$^4$λ, and c) z= 10$^5$λ.

High frequency spectral phase components will not be seen in this way, thus the main difficulty remains to determine the effects of spectral and spatiospectral phase distortions on the peak intensity.

Thus, measuring the pulse energy, the near field amplitude with cameras and imaging spectrometers gives the Fourier transform limit of the peak intensity. To estimate the real peak intensity, a full Strehl ratio estimation is then sufficient.

## III. Full Strehl ratio meter principle

The Strehl ratio, initially introduced in the spatial domain [25, 26], is the ratio of the peak intensity of a measured point spread function (PSF) to the peak intensity of a perfect diffraction limited PSF for the same optical system,

$$SR_{x_f y_f} = \frac{Max\,[I(x_f, y_f)]}{Max\,[I_{TF}(x_f, y_f)]} \quad (4)$$

In Adaptive Optics (AO), it is used for several purposes. One common usage, is to use the Strehl ratio maximization to optimize the AO corrections. In this usage, the absolute Strehl ratio does not matter so much as the relative changes in Strehl ratio between successive images. A Strehl ratio can also be used as a measure of system performance. As an example, it is related to the wave-front errors via the Maréchal approximations [27],

$$SR_{x_f} = \exp[-\sigma_\phi{}^2]\exp[-\sigma_\chi{}^2] \quad (5)$$

where $\sigma_\phi{}^2$ is the wave-front phase and $\sigma_\chi{}^2$ is the variance of the log-normal amplitude of the pupil plane. The Maréchal approximation shows that the total Strehl ratio is the product of the Strehl ratios of the individual components as long as the phase errors of each component are uncorrelated.

Similarly, for ultrashort pulses, spectral phase distortions also influence the compression of the pulse in the temporal domain. As the spectral phase and spatial ones are uncorrelated, it is possible to define a total Strehl ratio that also includes the temporal equivalent of the point spread function,

$$SR_{x_f y_f t} = SR_{x_f y_f} SR_t \quad (6)$$

The equality is valid only if the corresponding phases are uncorrelated. Spatiotemporal couplings will modify this relation.

Let's first consider that no spatiotemporal coupling affects the Strehl ratio and that all contributions are uncorrelated. Then the full Strehl ratio can be estimated from 1D estimations:

$$SR_{full} = SR_{x_f} SR_{y_f} SR_t \quad (7)$$

where $SR_{x_f} = \frac{\Delta x_{f_0}}{\Delta x_f}$, $\Delta x_f$ is the spatial width (in the sense of statistics i.e. the standard deviation) of the pulse and $\Delta x_{f_0}$ the spatial width of the FTL pulse (respectively $SR_{y_f}$ and $SR_t$ for orthogonal transverse dimension y and t for the temporal domain). Thus, measuring the Strehl ratio corresponds to measuring the broadening of the pulse in the spatial far field and temporal dimension:

$$SR_{full} = \frac{\Delta x_{f_0} \Delta y_{f_0} \Delta t_0}{\Delta x_f \Delta y_f \Delta t} \quad . \quad (8)$$

A great specificity of ultrashort pulses is the need [28, 29] but also the possibility of using nonlinear effects to characterize them. One of the latest methods proposed to measure ultrashort pulses [30] introduces a measurement quality criterion based on the spectral broadening of the pulse obtained by a third order nonlinear effect. This broadening is intrinsic to the instantaneous nonlinear effect. Let's consider that the pulse at its far field (or focal plane), goes through a nonlinear crystal that generates an nonlinear signal of order N either as harmonic or degenerate wavelength, as shown on figure 5. The input pulse is focused onto the nonlinear crystal and then the nonlinear signal is filtered out the input one by the filter. It is then collimated and measured.

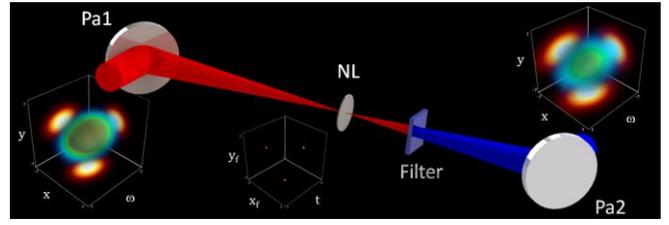

Fig. 5. Strehl ratio meter optical setup principle where NL is the nonlinear crystal, Pa1 and Pa 2 off axis focusing parabola.

The input pulse represented in the (x, y, $\omega$) domain is focused by the first parabola (Pa1) into the nonlinear crystal (NL). The nonlinear effect acts in the ($x_f$, $y_f$, t) domain and as long as the pulse is well compressed enough, it reduces the widths of the nonlinear pulse. After collimation of the nonlinear pulse by the second parabola (Pa2), the reduction effect directly by the broadening effect back in the (x, y, $\omega$) domain:

$$\Gamma_x = \frac{\Delta x_{NL}}{\Delta x}, \Gamma_y = \frac{\Delta y_{NL}}{\Delta y}, \Gamma_\omega = \frac{\Delta \omega_{NL}}{\Delta \omega}. \quad (9)$$

Indeed, the nonlinear effect broadening is maximal when the pulse is perfectly compressed i.e. for an FTL pulse. The relation between pulse broadening and spectral phase filtering has been studied for 3rd order nonlinear effects such as cross polarized wave generation (XPW) [31, 32] and results obtained for Gaussian pulses extend to any kind of pulses. If we consider a nonlinear effect of order N, and a gaussian pulse without any phase distortion at the input, then the pulse is reduced in the focal plane by the nonlinear effect by a factor $\sqrt{N}$. The bandwidth and the duration of a gaussian pulse are linked by: $\Delta \omega \Delta t \geq 1/2$ with the equality for FTL pulse. We then obtain for the broadenings:

$$\Gamma_{0x} = max(\Gamma_x) = \sqrt{N}, \; \Gamma_{0y} = \sqrt{N}, \Gamma_{0\omega} = \sqrt{N}. \quad (10)$$

As the relations are the same for x, y and $\omega$, let us detail only the relation according to $\omega$. We can wonder if the Strehl ratio, defined in the time domain is directly linked to the spectral broadening? If the pulse is stretched by a second order spectral phase, then its duration is given by:

$$\Delta t(\chi_2) = \sqrt{\Delta t_0{}^2 + \Delta t_G{}^2} = \Delta t_0\sqrt{1 + (\alpha_2 \chi_2)^2}. \quad (11)$$

While its spectral broadening is given by [32]:

$$\Gamma_{0\omega}(\chi_2) = \sqrt{\frac{1}{\Gamma_{0\omega}{}^2} \frac{\Gamma_{0\omega}{}^4 + (\alpha_2 \chi_2)^2}{1 + (\alpha_2 \chi_2)^2}} \quad (12)$$

where $\alpha_2 = \Delta \omega \Delta t_0$ and $\chi_2 = \frac{\varphi^{(2)}}{\Delta t_0{}^2}$ is the normalized chirp. One can observe that for a very large chirp and thus highly stretched pulse,

$$\Gamma_{\infty \omega} = \Gamma_\omega (\chi \rightarrow \infty) = \frac{1}{\Gamma_{0\omega}}. \quad (13)$$

So the spectral width is reduced by $\sqrt{N}$ in this case.

Finally, the Strehl ratio is expressed as a function of the spectral broadening:

$$R_{It}(\chi_2) = \frac{\Delta t_0}{\Delta t(\chi_2)} = \sqrt{\frac{\Gamma_\omega(\chi_2)^2 - \Gamma_{\infty \omega}{}^2}{\Gamma_{0\omega}{}^2 - \Gamma_{\infty \omega}{}^2}}. \quad (14)$$

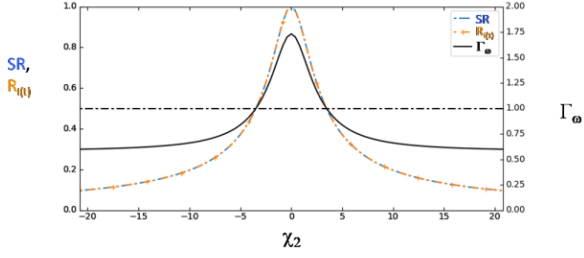

Fig. 6. Comparison of SR and $R_{I(\omega)}$ calculated from the measured broadening $\Gamma_\omega$ versus the normalized chirp $\chi_2$ for N=3.

For gaussian pulses, the fit is perfect when considering pure chirp (i.e. pure quadratic spectral phase). Physically, the reason is that the spectral broadening of the nonlinear pulse is indeed directly linked to the input pulse duration and its compression. If the spectral broadening is higher than 1 (the black curve is above the doted black one on fig.6), then the pulse duration is shorter than twice the duration of its Fourier limit i.e. the Strehl ratio is higher than 0.5 (c.f. fig.6).

The perfect fit over the entire range demonstrates that measuring the broadening gives a direct access to the Strehl ratio in this case.

For higher order spectral phase modulation, the pulse duration can still be expressed versus the normalized spectral coefficient of order n, $\chi_n$ as [32]:

$$\Delta t(\chi_n) = \sqrt{\Delta t_0^2 + \Delta t_g^2} = \Delta t_0^2 \sqrt{1 + (\alpha_n \chi_n)^2} \quad (14)$$

where $\alpha_n$ is a coefficient depending upon n and the shape of the spectrum. The relation remains the same. For the spectral broadening, as the pulse profile in time is modified and not only stretched as for a chirp, the broadening factor $\Gamma_\omega$ may vary with the phase distortion. This profile form factor can be estimated and corrected partially using the time bandwidth products of the input and the nonlinear pulse ($\Delta\omega\Delta t_0 = \alpha_1$), ($\Delta\omega_{NL0}\Delta t_{NL0} = \alpha_{NL}$). In any case, the estimation of the Strehl ratio by the broadening factor is better than 10% for broadening larger than 1.

If the spectral pulse profile is not gaussian, then $\Gamma_{0\omega} \neq \sqrt{N}, \Gamma_{\infty\omega} \neq 1/\Gamma_{0\omega}$, and $\alpha \geq \alpha_{NL} \geq 1/2$. As for spectral phases of order higher than 2, the estimation has an accuracy better than 10% when the broadening factor is larger than 1 and most of the discrepancy is due to the profile form that can be estimated via $\Gamma_{0\omega}, \Gamma_{\infty\omega}, \alpha_{NL}$ and $\alpha$. As an example, the limit case of a flat top temporal pulse has $\Gamma_{0\omega} \leq 1$ and thus is out of the scope of this article.

Intuitively, the quality of the approximation is understood by the direct link between pulse duration, spectral width and broadening factor.

This estimation in spectral domain can directly be transposed in the spatial domain. Thus, this method allows single-shot direct estimate of the 1D Strehl ratios $R_{I()}$ in every dimension $x_l$, $y_l$ and t, and if the phases are uncorrelated then the full Strehl ratio is given by:

$$R_{full\,I} = R_{Ix_f} R_{Iy_f} R_{It}$$
$$= \sqrt{\frac{\Gamma_x(\chi_{X2})^2 - \Gamma_{\infty X}{}^2}{\Gamma_{0X}{}^2 - \Gamma_{\infty X}{}^2}} \sqrt{\frac{\Gamma_Y(\chi_{Y2})^2 - \Gamma_{\infty Y}{}^2}{\Gamma_{0Y}{}^2 - \Gamma_{\infty Y}{}^2}} \sqrt{\frac{\Gamma_\omega(\chi_{\omega 2})^2 - \Gamma_{\infty\omega}{}^2}{\Gamma_{0\omega}{}^2 - \Gamma_{\infty\omega}{}^2}}. \quad (15)$$

As couplings in between dimensions can strongly affect the full Strehl ratio, are there 3D estimations possible with the same method including the couplings? As an example, the fig.7 presents the amplitude in the domain (x,y,t) (fig.7.d) of a pulse having a pulse front tilt in (x,t) that decreases its full Strehl ratio down to 0.1. The pulse front tilt is clearly shown by the image in the (x,t) domain where the pulse delay depends on the spatial x dimension (fig.7.f). The result in the (y,t) domain is that the pulse duration is strongly broadened (fig.7.a and b) while in the spatial domain (x,y) the pulse remains Fourier limited (fig.7.h-j).

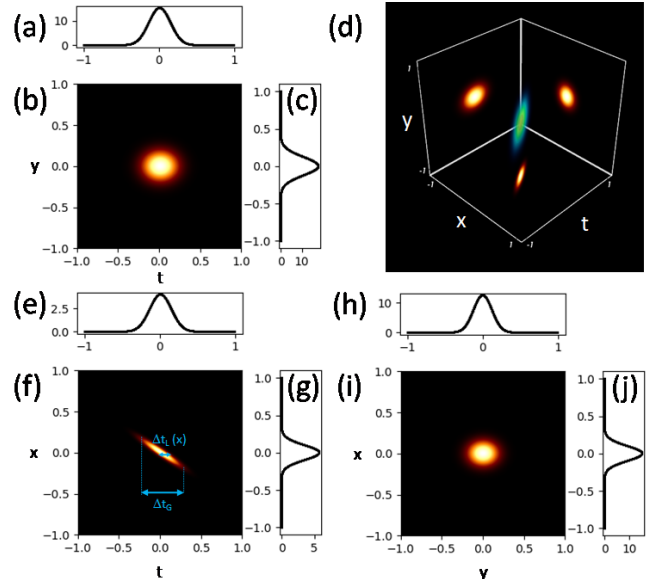

Fig. 7. Representations of intensities of PFT pulse in a,b,c) (t,y) domain with a) the integrated profile along y, b) the image of intensity over (t,y), c) the integrated profile along t, d) is the three dimensionnal representation of the amplitude of the pulse in (x,y,t) domain, with images as the integrated profile over one dimension (b,f,i), e,f,g) and h,i,j) are equivalent to a,b,c) but for (t,x) and (y,x) domains respectively.

The pulse front tilt in the (x,t) domain is perfect to illustrate the importance of global and local quantities [16] (fig.7.f). The local pulse duration is the FTL one as it does not integrate the pulse front tilt in x:

$$\Delta t_L(x,y) = \left[ \frac{\int_{-\infty}^{\infty} t^2 I(x,y,t)dt}{\int_{-\infty}^{\infty} I(x,y,t)dt} - \left( \frac{\int_{-\infty}^{\infty} tI(x,y,t)dt}{\int_{-\infty}^{\infty} I(x,y,t)dt} \right)^2 \right]^{1/2}. \quad (16)$$

While the global duration integrates all contributions, including the ones in x and y:

$$\Delta t_G = \left[ \frac{\iiint_{-\infty}^{\infty} t^2 I(x,y,t)dxdydt}{\iiint_{-\infty}^{\infty} I(x,y,t)dxdydt} - \left( \frac{\iiint_{-\infty}^{\infty} tI(x,y,t)dxdydt}{\iiint_{-\infty}^{\infty} I(x,y,t)dxdydt} \right)^2 \right]^{1/2}. \quad (17)$$

Most of the temporal intensity measurements are local ones and as shown on fig.7.f, measuring the local duration even at different positions do not allow determining the global duration.

The quantities that can be easily measured are spatial and spectral intensities. Thus, by using cameras and imaging spectrometers, the input pulse can be measured in the (x, y,ω) domain. The images obtained for the same pulse as the one of fig.7 are shown on fig.8 (b,f and i). As PFT is a pure spatiospectral phase effect, it cannot be seen on these measurements. Nevertheless, from these data, the FTL pulse can be estimated together with $\Gamma_{0\omega}, \Gamma_{\infty\omega}, \alpha_{NL}$ and $\alpha$.

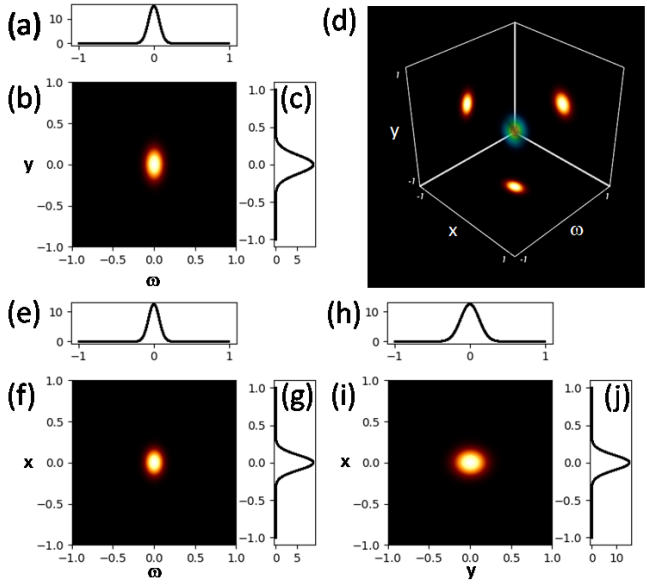

Fig. 8. Representations of spectral intensities of PFT pulse in a,b,c) ($\omega$,y) domain with a) the integrated profile along y, b) the image of intensity over ($\omega$,y), c) the integrated profile along $\omega$, d) is the three dimensional representation of the amplitude of the pulse in (x,y, $\omega$) domain, with images as the integrated profile over one dimension (b,f,i), e,f,g) and h,i,j) are equivalent to a,b,c) but for ($\omega$,x) and (y,x) domains respectively.

For this pulse that still has gaussian profiles, the values obtained for its FTL pulse are : $\Gamma_{0\omega} = 1/\Gamma_{0\omega a} = \sqrt{N}$, and $\alpha = \alpha_{NL} = 1/2$. No information on pulse duration or full Strehl ratio can be obtained from data of fig.8. As for 1D Strehl ratio meter method, the input pulse is focused into a nonlinear crystal. Fig.9 shows the output nonlinear results for a degenerated 3$^{rd}$ order nonlinear effect such as cross-polarized wave (XPW) [33,34] for example. As local and global pulse durations could be defined (fig.7.f), local and global pulse broadenings can be estimated from the measured amplitudes of fig.9. In theory, the broadenings can be on x, y or $\omega$ width function of other dimensions as for example $\Gamma_\omega$(x,y). Thus, it requires the full 3D measurement of fig.9.d. Unfortunately, only the integrated amplitudes and thus their relative broadenings are measured in single-shot with cameras and imaging spectrometers, as for example $\Gamma_{\omega(x)}$(y):

$$\Gamma_{\omega((x))}(y) = \frac{\Delta\omega_{NL(x)}(y)}{\Delta\omega_{(x)}(y)},\tag{18}$$

where
$$\Delta\omega_{NL(x)}(y) = \left[\frac{\int_{-\infty}^{\infty}\omega^2 I_{NL(x)}(y,\omega)d\omega}{\int_{-\infty}^{\infty} I_{NL(x)}(y,\omega)d\omega} - \left(\frac{\int_{-\infty}^{\infty}\omega I_{NL(x)}(y,\omega)d\omega}{\int_{-\infty}^{\infty} I_{NL(x)}(y,\omega)d\omega}\right)^2\right]^{1/2}$$

and $I_{NL(x)}(y,\omega) = \int_{-\infty}^{\infty} I_{NL}(x,y,\omega)dx$ for the nonlinear pulse (NL) and for input pulse by replacing NL subscript by 0. It simplifies by considering centered and normalized intensities:

$$\Delta\omega_{NL(x)}(y) = \left[\int_{-\infty}^{\infty}(\omega - \omega_0)^2 I_{NL(x)}(y,\omega)d\omega\right]^{1/2}$$
$$= \left[\int_{-\infty}^{\infty} \Delta\omega_{NL}{}^2(x,y)dx\right]^{1/2}.\tag{19}$$

Equation (19) means that the spectral width measured on the image ($\omega$,x) is the global one and corresponds to the sum of the local ones with respect to x. By analogy on the other dimensions, the global widths are the one obtained through the integration over the other dimension, i.e. on the three images $\Gamma_{\omega<y>}$(x)and $\Gamma_{x<y>}(\omega)$on ($\omega$,x), $\Gamma_{\omega<x>}$(y)and $\Gamma_{y<x>}(\omega)$ on ($\omega$,y) and $\Gamma_{y(\omega)}$(x) and $\Gamma_{x(\omega)}$(y) on (x,y). Experimentally, the estimation of the width is very sensitive to any noise, background and offset. So the broadening should be determined preferentially by another method.

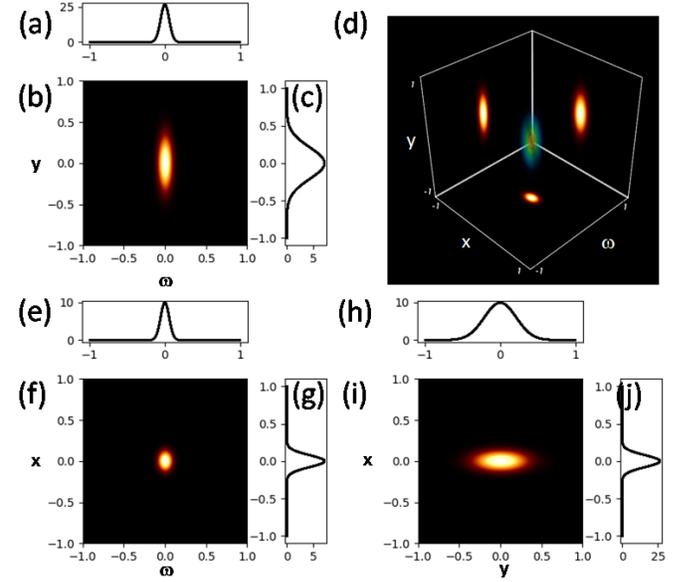

Fig. 9. Representations of intensities of 3$^{rd}$ order nonlinear PFT pulse in a,b,c) ($\omega$,y) domain with a) the integrated profile along y, b) the image of intensity over ($\omega$,y), c) the integrated profile along $\omega$, d) is the three dimensional representation of the amplitude of the pulse in (x,y, $\omega$) domain, with images as the integrated profile over one dimension (b,f,i), e,f,g) and h,i,j) are equivalent to a,b,c) but for ($\omega$,x) and (y,x) domains respectively.

By analogy with gaussian pulses [32], a gaussian pulse equivalent broadening can be estimated as the inverse of the ratio of the maximum of the normalized spectral intensity over the maximum of the nonlinear pulse of Fourier transform limited input one.

Thus, from these three different curves and the three images, six broadenings that characterize the pulse and the Strehl ratio can be estimated. As an example on the image of fig.9.f, one can estimate the broadening as:

$$\frac{\Gamma mes_{\omega x(y)}}{\Gamma_0} = \left\{\frac{\max\left(I_{NL(y)}(x,\omega)\right)}{\int I_{NL(y)}(x,\omega)dxd\omega} \bigg/ \frac{\max\left(I_{TFNL(y)}(x,\omega)\right)}{\int I_{TFNL(y)}(x,\omega)dxd\omega}\right\},\tag{20}$$

where $I_{TFNL<y>}$ is the spectral intensity of the calculated nonlinear pulse obtained from the FTL input one, integrated over y.

The same kind of broadening estimations $\frac{\Gamma mes_{\omega y(x)}}{\Gamma_0}$ and $\frac{\Gamma mes_{xy(\omega)}}{\Gamma_0}$ can be done on images of fig.9.a and d, or from the three curves, the estimations $\frac{\Gamma mes_{\omega(y)x}}{\Gamma_0}$, $\frac{\Gamma mes_{\omega(y)x}}{\Gamma_0}$ and $\frac{\Gamma mes_{(x)\omega y}}{\Gamma_0}$. From the measured full broadening $\Gamma mes_{\omega(xy)}$ that corresponds to the measured full Strehl ratio, $R_{Imes\omega<xy>}$ can be estimated as:

$$R_{Imes\,\omega(xy)} = \sqrt{\frac{N^2}{N^2-1}\left(\left(\frac{\Gamma mes_{\omega(xy)}}{\Gamma_0}\right)^2 - \left(\frac{1}{N^2}\right)\right)}.\tag{21}$$

Thus six estimations of the Strehl ratio are done in different domains by projection of the intensity. The table 3 below sums up the different domains and gives some example of results for basic distortions such as chirp, PFT, PFC, TGDD. These estimations are compared to the full Strehl ratio calculated on the full 3D data in these simulated examples. As shown by these few examples, the measures are close to or at least a bit higherthe full Strehl ratio even for relatively large distortions. It is possible to identify in which domains the phase distortion happens. Indeed for a pure chirp in optical frequency, all the pure spatial domains are perfectly compressed and give an equivalent Strehl ratio equal to 1. While for STC like PFTx that couples x and $\omega$, only the 1D

Strehl ratio estimation over y is equal to unity.Higher order STCs like PFC or TGDD are also properly estimated and their dimensions easily identified. Even the consequence of a combination of STCs like PFT and TGDD is precisely determined on the full Strehl ratio.

|  | Domain | FTL | Chirp ω | PFTx | PFCy | TGDDx | PFTy + TGDDy |
|---|---|---|---|---|---|---|---|
| **SR_full** | **3D:(x,y, ω)** | **1** | **0.1 0.5** | **0.1 0.5** | **0.5** | **0.5** | **0.5** |
| $R_{Imesωx}$ <y> | 2D:(x,ω) | 1 | 0.16 0.5 | NaN 0.47 | 0.81 | 0.49 | 0.72 |
| $R_{Imesωy}$<x> | 2D:(y,ω) | 1 | 0.16 0.5 | 0.33 0.71 | 0.5 | 0.62 | 0.5 |
| $R_{Imesxy}$<ω> | 2D:(x,y) | 1 | 1 1 | 0.32 0.7 | 0.62 | 0.81 | 0.74 |
| **$R_{I2D}$** | **Min(1D)** | **1** | **0.16 0.5** | **NaN 0.47** | **0.5** | **0.49** | **0.5** |
| $R_{Imes3D<ω>xy}$ | 1D:(ω) | 1 | 0.16 0.5 | 0.33 0.71 | 0.81 | 0.62 | 0.72 |
| $R_{Imes3D<x>ωy}$ | 1D:(x) | 1 | 1 1 | 0.32 0.7 | 1 | 0.81 | 1 |
| $R_{Imes3D<y>ωx}$ | 1D:(y) | 1 | 1 1 | 1 1 | 0.62 | 1 | 0.74 |
| **$R_{I3D}$** | **$1D^3$** | **1** | **0.16 0.5** | **0.1 0.5** | **0.5** | **0.5** | **0.53** |

TABLE 3Strehl ratios comparison from theoretical one to measured ones on the different domains, versus no distortion, pure spectral chirp, PFT over x, PFC over y, TGDD over x and a mix of PFT and TGDD along y.

A simple combination of 1D or 2D estimations by multiplication, gives estimated Strehl ratios:

$$R_{I2D} = R_{Imes\ ωx(y)} R_{Imes\ ωy(x)} R_{Imesxy\ (ω)} ,\quad (22)$$

$$R_{I3D} = R_{Imes\ ωx(y)} R_{Imesx\ (ωy)} R_{Imesy\ (ωx)}.\quad (23)$$

These estimations are very close to the full Strehl ratio even for complex phase distortions like a combination of PFT and TGDD along y.

As expected from the broadening factor sensitivity [32], the precision of the measure is good when the broadening is larger than 1 i.e. for Strehl ratio close to 0.5 or higher. Larger values can be estimated but the precision of the measurement decreases as the broadening decreases and gets close to $\Gamma_\infty$.

As it is not possible to scan over all the phase distortions in all dimensions, to illustrate the ability of the Strehl ratio meter method, the results of the simulations are determined through a map of values over distortions of 2nd and 3rd order in two dimensions (x,ω). For the 2nd order, a combination of chirp and PFT that induces distortion with a Strehl ratio decreasing down to 0.1 is well measured by the method with an absolute error below 10% as shown on fig.10.a where the 3D iso-surfaces represent the Strehl ratio while the absolute error in percent defined as $-100*(SR_{full}-R_{I3D})/max(SR_{full})$ is represented as a vertical vector (along divergence direction) at each point whose color is coded over its scalar value from black to red for 100%. The inset fig.10.a illustrates that the error is lower than 10% when the Strehl ratio is higher than 0.5 i.e. when the nonlinear pulse is spectrally broadened. The estimation is still rather good even for a ratio down to 0.1 as shown on fig.10.b. The maximum values of the error are on the side where the Strehl ratio was as low as 0.02. On fig.10, the error is always positive meaning that the Strehl ratio is over estimated by the measurement. This over estimation, meaning that the measured Strehl ratio is always higher than the real one, is mainly due to its estimation of global parameters. The full Strehl ratio is a local maximum that corresponds to the ratio of the energy over the local width. It is always higher than the maximum of the integrated intensity over one or two dimensions. Indeed except for example if the maximum position over t is the same for each position (x,y), the global maximum max($I_{<xy>}$(t)) is lower than the local one max(I(x,y,t)). It also corresponds to the fact that the global widths are always larger by definition than local ones. To further confirm the quality of our Strehl ratio meter, the Strehl ratio was simulated over third order distortions (Phi(3), TGDDx and PFCx). The fig.11 illustrates the complexity that can arise from third order distortions. Despite this complexity, the error is still lower than 10% over a range down to Strehl ratio as low as 0.25 as shown on fig.11.a. The iso-surface in green in fig.11.b corresponds to Strehl ratio level of 0.25 and illustrates the complexity of the evolution of the Strehl ratio over such distortions. The error is still always positive and can be up to 75% on such complex distortions for a Strehl ratio down to 0.07.

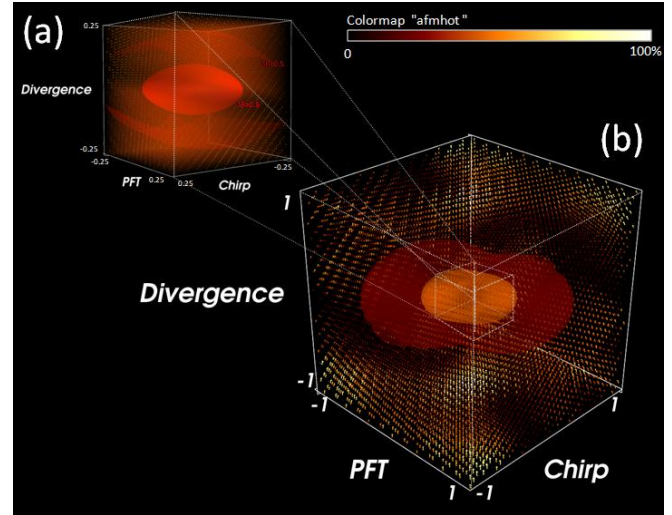

Fig. 10. Representations of full Strehl ratio as iso-surfaces (0.5 and 0.8 for (a), 0.25 and 0.5 for (b)) versus the chirp, PFT and divergence over x. (a) is the inset of (b). The vector fields represent the absolute error of the measurement versus the Strehl ratio over a "afmhot" colormap.

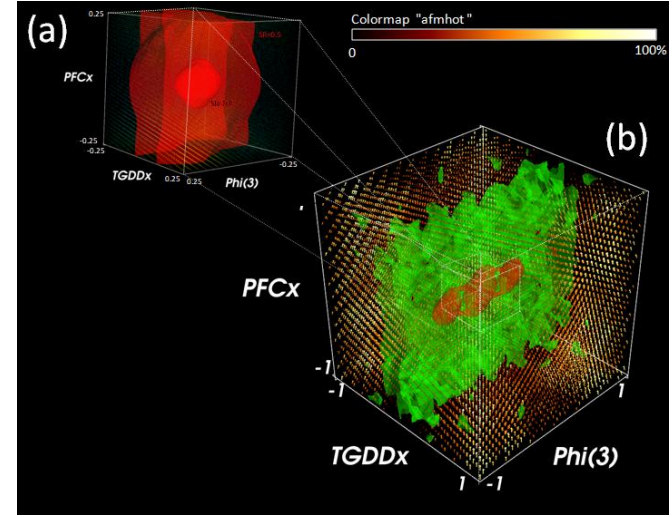

Fig. 11. Representations of full Strehl ratio as iso-surfaces (0.5 and 0.8 for (a), 0.25 and 0.5 for (b)) versus third orders distortions Phi(3), TGDDx and PFCx over x. (a) is the inset of (b). The vector fields represent the absolute error of the measurement versus the Strehl ratio over a "afmhot" colormap.

This illustrates that this method is very efficient and precise for pulses that are not too strongly distorted. But despite a rather low precision when the pulse is strongly distorted by third orders, the result still proves that there is some distortions that affects the pulse. In addition, when combined with the effectiveness of the non-linear effect, it undoubtedly characterizes the poor compression of the pulse. To further consider more complex phase distortion consequences, it is important to explore also higher frequency phase distortion components and check if it still diagnoses the "illness" of the pulse.

## IV. Phase distortion frequency considerations

In the spatial domain, to evaluate the influence of the phase distortions induced by surface by surface quality of mirrors or lenses, spatial frequencies are used to differentiate between RMS gradient, waviness or surface roughness [35]. The influence of the different components has been widely studied for NIF project [36-38]. The range of frequency is divided into three separate bands. The wavefront quality over the first band of low frequencies (2.5x10-3 to 3.0x10-2 mm-1) or large spatial scales (400 to 33 mm) is specified in terms of a maximum peak-to-valley error (<λ/3) and a maximum rms phase gradient (<λ/90/m measured at 633nm). This first domain is also defined as the low-frequency "form" errors [35]. The second spatial frequency region (3.0x10-2 to 8.3mm-1) is specified through a maximum "waviness" in terms of power spectral density (PSD). The PSD simply represents the square of the phase noise amplitude (nm²) over a certain spatial frequency (mm-1). This spatial frequency domain is also named Mid-Spatial Frequency domain. The third spatial frequency band is from 8.3 to 100mm-1 and represents the micro-roughness region. As expected by Fourier optics [20] the focal spot shape depends mainly on the low frequency part, but the encircled energy or the Strehl ratio is indeed affected by all the domains [39,40]. The exact same effect takes place in the temporal domain where spectral phase frequency induces pulse distortions for low frequency spectral phase distortions while the temporal contrast on ps and ns scale is due to mid and high frequency components as illustrated on fig.12.Three different phase frequency distributions are applied in x, y and ω domains. They all result in a Strehl ratio around 0.1 with a very good estimation from our method. Indeed, the important parameter for the pulse distortion is the gradient of the phase along ω for the temporal domain and x and y for the spatial ones. Thus for a same amplitude, a high frequency component will cause an energy spread farther from the central point. It also explains difficulties of a direct relation between phase frequency components and pulse distortions.

Historically, the different domains were issued from the available metrology resource for surface characterization [37,38] while the different domains should be linked to their result in the image plane. Where in these frequency domains can the Strehl ratio meter method introduced here be placed?

Without metrology considerations how to estimate low, mid and high spatial frequency domains?

Aikens et al. [39] defines a different domain. The low spatial frequency (LSF) is the domain described by Zernike fit of the aberrations up to the 37 term, i.e. 5-10 cycles per aperture. The mid-spatial frequency (MSF) upper limit is the boundary where roughness becomes predominant, i.e. when the Fresnel number is lower than 0.1, or when the spatial frequency is higher than (λxOPL/10)^-1/2 where OPL is the Optical Path Length from the phase distortion to the image plane. The high spatial frequency (HSF) part is essentially controlled with an RMS roughness specification that leads to losses but without distortion of the pulse. Spatially, the scattering of the light is too important to reach the image plane. The ultimate upper limit for the spatial frequency is twice the inverse of the wavelength. In our case we consider that the near field has an aperture a=10cm. So the MSF starts at (10/a)=0.1mm-1. The upper limit depends on the Optical Path Length. For the spectrotemporal domain, the OPL is infinitely long. Thus, there is no upper limit and no high-spatial frequency domain. For the spatial domain, we can consider at least that the OPL is the distance from the last optic to the focal spot. On high power lasers, ideally the last optic is a parabola with a maximal aperture i.e. a=OPL. Thus the HSF limit is about:

(aλ/10)^-1/2≈10mm-1.

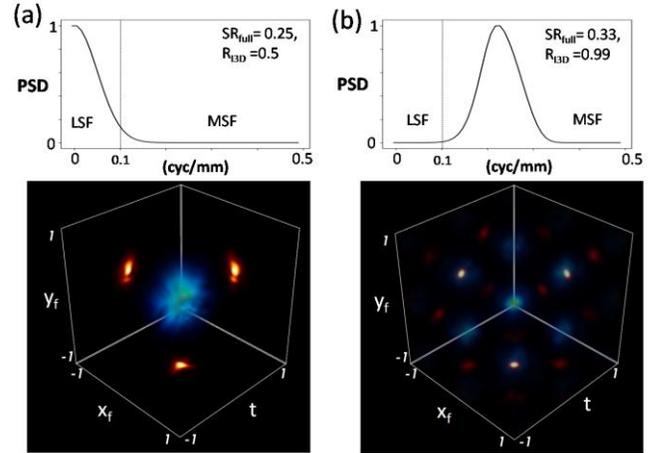

Fig.12: illustration of (a) low and (b) mid spatial frequency regions influences onto the focal spot for Spectral phase with PSD.

As shown by the fig.12, if the phase frequency distribution is mainly in the low-spatial frequency domain (LSF), the method is efficient to recover the Strehl ratio as long as the main pulse is affected. The low frequency part represented in fig12.a, where most of the power spectral density is close to zero, induces a pulse distortion in $x_f$, $y_f$ and time domains that decreases the Strehl ratio down to 0.25. The estimation recovered by our method is about 0.5 and is even closer if we use the spectral width (0.25). Identically, fig.12.b shows the same characteristics respectively for mid-frequency components. The few distortions visible on a linear scale spread most of the energy out of the main pulse with a Strehl ratio around 0.33. As the main pulse is not affected, our method overestimates the Strehl ratio (0.99). Using the spectral width and the Fourier transform limited pulse duration allows getting a closer estimation (0.46). Interestingly the difference between these two estimations gives an insight on the origin of the distortion. For the higher spatial frequency domain, our 3D simulation is not able to manage the high frequency part in full 3D. As shown on fig.4, if the spatial frequencies are high enough, the amplitude profile is strongly modified after a short propagation compared to the OPL (Fresnel number=(a²/λxOPL) >1). Thus, a simple diagnostic with a camera fulfilled the need for this spatial frequency part. On the spectral domain, there is no high-spatial frequency because the equivalent OPL is infinite. Our method despite its quantitative weakness on this part is still able to diagnose a contrast pathology due to these phase components as shown on fig.13. The phase distortions are only applied to the spectral domain while the spatial dimensions have no distortion. As a result, the focal spot intensity is only affected in the time domain. The fig.13.a-e present spectral phase with frequencies in the beginning of the mid-spatial frequency area equivalent from 0.2 to 2.5mm-1 with a decrease of the Strehl ratio down to SR_full=0.22. This energy loss in the focal spot is not seen by our method (R_{3D}=0.9). By using the spectral profile and determining its width and the duration of its Fourier limit equivalent pulse for the initial and nonlinear pulses, an equivalent estimation of the Strehl ratio gives another estimation which is closer in this case (R_{1Δω} = 0.01). It is not a quantitative estimation but as for lower frequencies of fig.12 it is an indicator that mid-spatial frequencies do distort the pulse. Even for higher frequencies like on fig.13.f-j that strongly spread 90% of the energy out of the main pulse, this other indicator gives a nearly null result (R_{1Δω} = 0.01) that clearly highlights the problem. Nevertheless, this indicator is more affected by experimental noise as it relies on statistical width estimations:

$$R_{1Δω} = \min(1, \alpha/\alpha_{NL}) \sqrt{\left| \frac{\Gamma mes_{Δω}^2 - \Gamma_{\infty\omega}^2}{\Gamma_{2\omega}^2 - \Gamma_{\infty\omega}^2} \right|}, \qquad (24)$$

where $\Gamma mes_{Δω} = \Gamma_{\omega(xy)} = \frac{Δω_{NL(xy)}}{Δω_{(xy)}}$.

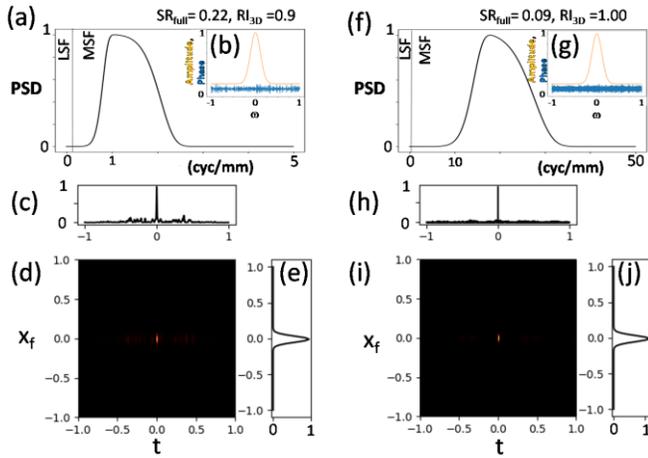

Fig.13: illustrations of mid spatial frequencies on spectral phase only influence onto the focal spot, a-e) for few cycles/mm and f-j) few 10s cycles/mm. (a,f) normalized PSD of the spectral phase. (b,g) respective amplitude (orange curve) and phase (blue curve) of the pulses considered. (c,h) temporal amplitude profiles. (d,i) far field spatiotemporal amplitude integrated over $y_f$. (e,j) spatial amplitude profiles.

Thus it acts more like a diagnostic rather than a metrology tool. As symptoms do not clearly indicate the cause of an illness, it shows on a single-shot acquisition that the pulse is not in a good shape. As for a patient in medicine, after a first diagnosis of illness, further investigations are needed to identify the cause of the illness and/or to cure it.

## V. Diagnostic of phase distortions causes

As it is not possible in a single article to fully describe all distortions' causes for a pulse, we will focus on main already known defaults of CPA laser chains. A first method introduced by Margoto et al. [41] shows the ability of this method to identify PFT, TGDD and PFC on the pulse. The principle of the method is to diagnose defaults on the pulse compression and then to find their origin by complementary measurements with additional chirp or divergence. Depending upon the phase distortions origin, the complementary measurements may differ. In the article of Margoto et al., the authors use chirp to differentiate the Strehl ratio decrease caused by a combination of chirp and divergence or by a Pulse front tilt. The analysis is done by comparing the spatiospectral image of the nonlinear pulse versus the input pulse.

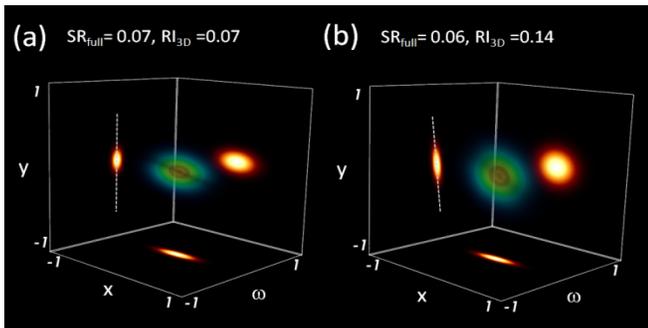

Fig.14: a) Nonlinear pulse spatiospectral intensity with PFT not appearing, b) Nonlinear pulse spatiospectral intensity with additional chirp that reveals PFT through central frequency evolution versus Y.

Two parameters, the central position and the broadening clearly indicate the origin of the distortion. The fig.14.a shows the nonlinear pulse in (x,y,ω) domain. Its initial pulse has a pulse front tilt in (x,ω) domain that does not appear clearly on fig.14.a. By adding a chirp to the initial pulse, the PFT is revealed by a linear dependence of the central position of the spectrum versus y as highlighted by the dotted line tilt in fig.14.b compared to the one in fig.14.a.

To exemplify more clearly the problem, fig.15 graphically sums up the way to diagnose a PFT problem in (x,ω) domain on a CPA laser chain. Starting from the initial pulse amplitudes in (x,ω) domain, its Fourier limit pulse nonlinear spatiospectral image can be calculated (fig.15.a). From this image, the central positions (Δω versus x and Δx versus ω) are calculated and superimposed by blue lines directly onto the image. The broadenings ($\Gamma_\omega(x)$ and $\Gamma_x(\omega)$) are plotted on each respective side of the image (a.1 and a.2) and the obtained curve (blue) is compared to the reference level 1 represented as a dotted line. If the pulse is perfectly compressed, the measured nonlinear spatiospectral amplitude should be the same. The central positions are along strait lines, vertical for the optical frequency versus x and horizontal for the central position versus ω. As expected, for a perfectly compressed pulse, the broadenings are maximum, equal to $Z_0$. Adding a significant chirp that decreases the Strehl ratio down to 0.47 affects only the broadening of the pulse in the frequency domain i.e. the broadening curve on the right side $\Gamma_\omega$ versus x. By symmetry, adding a divergence of the same amount decreases the spatial broadening versus optical frequency down to 1 and the Strehl ratio also down to 0.47 (fig.15.c). The central positions are not affected and the broadening in the other dimension remains maximal. If now a combination of this same chirp and divergence are added to the initial FTL pulse, the Strehl ratio is decreased down to 0.22 and the broadenings reveal that both dimensions are affected (fig.15.d). Ambiguously, a PFT can exactly introduce the same distortions as shown on fig.15.g. Thus the measurement gives the Strehl ratio with an excellent precision but cannot distinguish between PFT or chirp and divergence combination. To identify the origin of the distortion, additional measurements are needed.

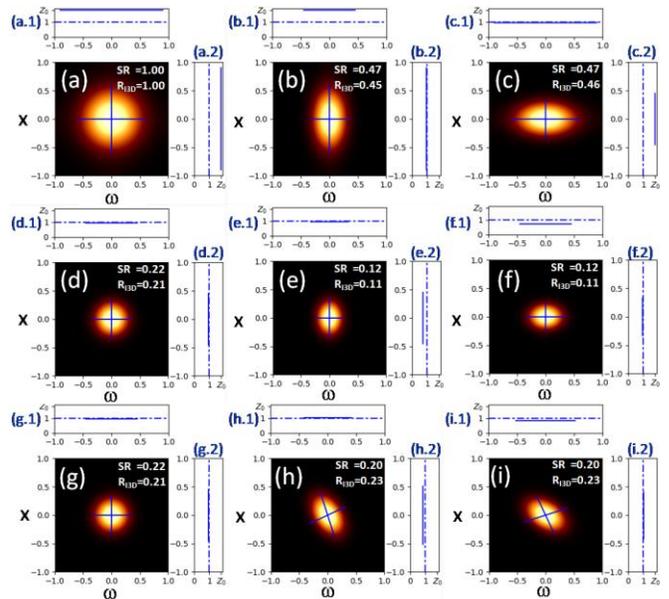

Fig.15: Nonlinear pulse spatiospectral intensities in (x,ω) domain for a Fourier Transform limited pulse (a), the blue lines on the image are the central positions, and the curves on the side (a.1 and a.2) are the local broadenings, (b) and (c) is the same data for the pulse of (a) with additional chirp and divergence respectively. (d) is for a pulse with chirp and divergence,(g) for a pulse with PFT (e), (h) and (f),(i) respectively with additional chirp and divergence.

If we add to the pulse the same amount of chirp which was introduced in fig.15.b, the pulse with chirp and divergence has its Strehl ratio decreased again and only the spectral broadening is affected (fig.15.e). For the pulse with a PFT, the same chirp will introduce on the nonlinear spatiospectral image a totally different deformation (fig.15.h). The Strehl ratio is nearly not affected while the central positions are tilted, the straight lines remain straightt but their directions are tilted from horizontal and vertical. These tilts are characteristics of a PFT. By symmetry, the divergence will also introduce the same effects (fig.15.i). It perfectly illustrates the way to diagnose the origin of the distortion of the second order with our method. The central positions and broadenings give a clear insight of the pulse distortions and help to diagnose its origins. Only a PFT can introduce this tilt. The value of the PFT may be evaluated by simulation and comparison with the initial data using blind algorithm for phase retrieval. As the data of fig.15 are 2D and redundant, the convergence of the algorithm should be efficient [42]. Nevertheless, our main interest is in the single-shot or few shots diagnostic capability. The sensitivity of the method is excellent even for off-the-shelf imaging spectrometers. More generally, spatiospectral image deformation is very typical of STC effects, analogous to the image distortions seen in frequency resolved optical gating traces due to spectral phase defects [43].

The fig.16 extends this example to the third order STCs, the PFC and the TGDD. The PFC with a significant effect on the pulse (SR=0.35 in fig.16.a) will be better revealed by the curvature of the central position by a divergence (fig.16.c) rather than a chirp (fig.16.b). By symmetry, the TGDD introducing about the same distortion on the pulse (fig.16.d) is identified more easily by the curvature on the central frequency introduced by a chirp rather than by a divergence. It is even clearer on larger TGDD (fig.16.g, h, i).

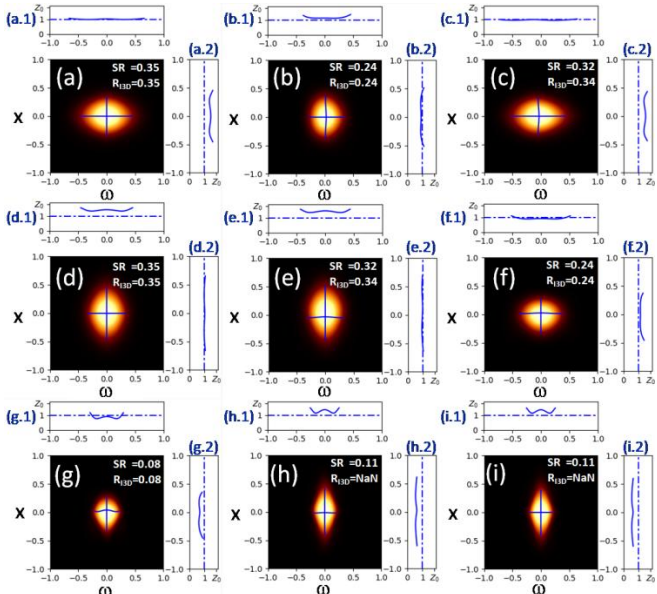

Fig.16: Nonlinear pulse spatiospectral intensities in $(x, \omega)$ domain for third order STCs (a) PFC, (d) and (g) TGDD, the blue lines on the image are the central positions, and the curves on the side are the local broadenings, (b), (e), (h) and (c), (f), (i) are the same data with additional chirp and divergence respectively.

The curvature is even clearer on the broadening curves. The curvature is due to the quadratic term ($x^2$ for the PFC and $\omega^2$ for the TGDD). As already shown in the previous part, the Strehl ratio estimation lacks precision when the distortions are large or when the spatial equivalent frequencies are too high. Another estimator $R_{i\Delta\omega}$ should then be used. But the ability to diagnose the distortions' origins is still valid even for more general distortions. The fig.17 exemplifies one case for LSF (a,b,c) and two MSF cases whose phase modulations amplitude is large for d,e and f and 0.7 smaller for g, h and i. For LSF, the diagnostic is exactly as simple as it is for second and third order. Central positions and frequencies show up the distortions and their origin (strong STCs). The Strehl ratios are better estimated by $R_{i\Delta\omega}$ but lack from precision despite revealing clues on central positions and broadenings. New estimators with a better precision should be possible eventually including new parameters or blind algorithms but already these images allow a very intuitive interpretation of the faults present.

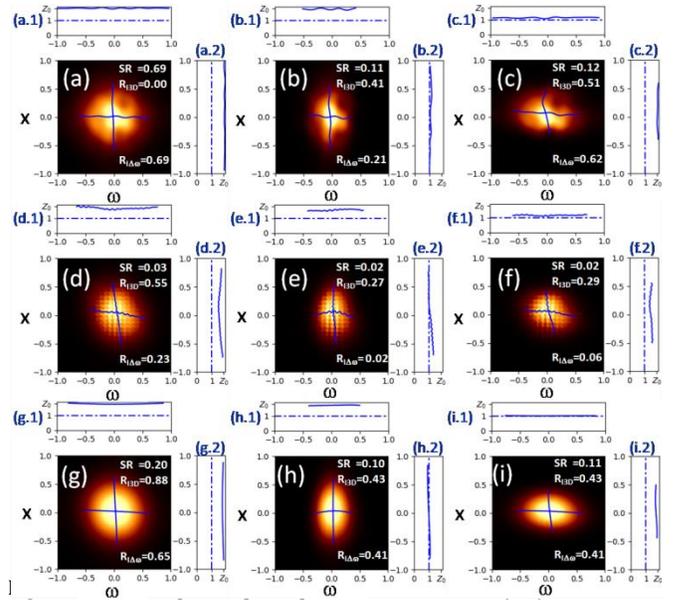

(a) LSF distortions, (d) MSF large distortions or (g) MSF lower ones , the blue lines on the image are the central positions, and the curves on the side are the local broadenings, (b), (e), (h) and (c), (f), (i) are the same data with additional chirp and divergence respectively.

## VI. Conclusion

The article presents the theoretical part of a newly introduced method [41] and extends it to any other nonlinear effects and to the Strehl ratio metrology. As a single-shot diagnostic, it will benefit any large laser facility and, by its simplicity, will allow any user to check the pulse quality on every shot. In the same way as the thermometer, by measuring the body temperature, is the first diagnosis of a disease, the ultrashort pulse Strehl ratio meter is the first clue in the diagnostic of a pulse distortion. If the estimated ratio is 1 and the broadenings are maximum ($\Gamma_{0x}$, $\Gamma_{0y}$ and $\Gamma_{0\omega}$), then the pulse is in perfect shape. If it is not, then additional measurements are required to diagnose the origin of this illness. Some can directly be made with the same method by adding chirp or divergence. This already covers a large range of known faults of CPA lasers, but others may require new diagnostics. This is the first method in the ultrashort laser community that introduces a methodology similar to diagnosing a large complex system like a human body. Its unambiguous results and its simplicity of use will make it "the thermometer" of the high-intensity ultrashort laser community.

A first experimental validation of the method was carried out by [41] demonstrating its ability to diagnose a complex STCs distortion with a good sensitivity. Other implementations are possible because the method can be applied to any "instantaneous" nonlinear effect. In particular, surface harmonic generation that was used in [44] could be used to characterize in situ at the focal spot the full Strehl ratio on the target. This would thus fulfill the essential need for determining the peak intensity on every shot as it opens the way for full single shot diagnosis of focus pulse on target.

**Acknowledgment**. I would like to thank Dr. Lopez-Martens, Dr. Jean-Baptiste Margoto, Dr. Olivier Gobert, Mathias Herzog for fruitful discussion and attentive reading of this manuscript. I also thank Mayavi [45], MatplotLib [46], Scipy [47] and NumPy [48, 49], and Python developers.

**Disclosures.** The authors declare no conflicts of interest